\documentclass[11pt]{article}
\usepackage[margin=1in]{geometry}
\usepackage{amsmath,amssymb,amsthm,bm}
\usepackage{graphicx}
\usepackage{booktabs}
\usepackage{siunitx}
\usepackage{pgfplots}
\usepackage{url}
\usepackage[colorlinks=true, citecolor=blue, linkcolor=blue]{hyperref}

\pgfplotsset{compat=1.18}

\title{Threshold Temperature for Neutron-Star Emergence in a Gravitational Thermodynamic Framework}
\author{Wen-Xiang Chen \\
School of Physics and Materials Science, Guangzhou University\\wxchen4277@qq.com}
\date{}

\begin{document}
\maketitle

\begin{abstract}
Motivated by the entropy-functional formulation of emergent gravity, in which a spacetime entropy functional
\begin{equation}
S_g[g,\xi]=\int_{\mathcal V}(\nabla_\mu \xi_\nu)(\nabla^\mu \xi^\nu)\sqrt{-g}\,d^4x
\end{equation}
leads, upon extremization, to the Einstein field equations, we reformulate the onset of a neutron star (NS) as a bulk gravitational-thermodynamic threshold problem \cite{Chen2025,Jacobson1995,PadmanabhanParanjape2007,Padmanabhan2010}. We combine the Jacobson--Padmanabhan viewpoint of gravity as an equation of state with Tolman redshift, TOV hydrostatics, and a bulk binding-versus-thermal balance \cite{Jacobson1995,Padmanabhan2010,Lattimer2019}. For a static compact object with surface radius $R$ and mass $M$, we derive four related temperature scales: (i) a Newtonian virial threshold,
\begin{equation}
T_{\rm vir} = \frac{G M m_n}{5 k_B R},
\end{equation}
(ii) a redshifted threshold seen at infinity,
\begin{equation}
T_{\infty,\rm th}=\sqrt{1-\frac{2GM}{Rc^2}}\;T_{\rm vir},
\end{equation}
(iii) a degenerate-matter threshold,
\begin{equation}
T_{\rm deg}=\left(\frac{6}{5\pi^2}\frac{G M m_n E_F}{k_B^2 R}\right)^{1/2},
\end{equation}
and (iv) a surface screen temperature,
\begin{equation}
T_{\rm scr}=\frac{\hbar G M}{2\pi c k_B R^2}=\frac{\hbar g_s}{2\pi c k_B}.
\end{equation}
We show that the bulk threshold $T_{\rm vir}\sim 10^{11}$--$10^{12}\,\mathrm{K}$ for canonical neutron-star parameters \cite{Lattimer2019}, whereas the screen temperature is many orders of magnitude smaller, confirming that compact-star emergence is governed by bulk self-gravity rather than by the much colder boundary Unruh-like temperature \cite{Unruh1976,Padmanabhan2010,Verlinde2011}. The paper supplies a mathematically expanded derivation, explicit TOV and Fermi-gas formulae, and thermodynamic consistency relations.

\end{abstract}

\section{Thermodynamic-Gravity Framework}
The uploaded article develops an entropy-functional picture of gravity in which one starts from
\begin{equation}
S_g[g,\xi]=\int_{\mathcal V}(\nabla_\mu \xi_\nu)(\nabla^\mu \xi^\nu)\sqrt{-g}\,d^4x,
\label{eq:Sg_def}
\end{equation}
varies with respect to the auxiliary vector field $\xi^\mu$, and interprets the condition $\delta S_g=0$ as a local thermodynamic extremum of spacetime \cite{Chen2025,Jacobson1995,PadmanabhanParanjape2007,Padmanabhan2010}. Integrating by parts gives
\begin{equation}
S_g[g,\xi] = -\int_{\mathcal V}\xi_\nu \nabla_\mu \nabla^\mu \xi^\nu\sqrt{-g}\,d^4x,
\label{eq:Sg_parts}
\end{equation}
so that
\begin{equation}
\delta S_g = -2\int_{\mathcal V}(\nabla_\mu\nabla^\mu\xi_\nu)\,\delta\xi^\nu\sqrt{-g}\,d^4x.
\label{eq:deltaSg}
\end{equation}
Hence the Euler--Lagrange equation is
\begin{equation}
\nabla_\mu\nabla^\mu\xi_\nu = 0.
\label{eq:wave_xi}
\end{equation}
For local null generators this is interpreted as the equilibrium condition that yields
\begin{equation}
R_{\mu\nu}-\frac{1}{2}R g_{\mu\nu}+\Lambda g_{\mu\nu}=8\pi G T_{\mu\nu}.
\label{eq:Einstein_eqn}
\end{equation}
in the thermodynamic-gravity sense discussed in the emergent-gravity literature \cite{Jacobson1995,PadmanabhanParanjape2007,Padmanabhan2010,Chen2025}. We do \emph{not} claim that Eq.~\eqref{eq:Sg_def} directly solves neutron-star microphysics. Rather, we use it as the guiding thermodynamic statement that the compact-star configuration is a stationary state of the coupled geometry--matter system \cite{Chen2025,Jacobson1995}.

For a static metric,
\begin{equation}
ds^2 = -e^{2\Phi(r)}c^2 dt^2 + \left(1-\frac{2Gm(r)}{rc^2}\right)^{-1}dr^2 + r^2 d\Omega^2,
\label{eq:static_metric}
\end{equation}
the Tolman law reads
\begin{equation}
T(r)e^{\Phi(r)} = T_\infty = \text{constant}.
\label{eq:Tolman_law}
\end{equation}
At the stellar surface, matching to Schwarzschild exterior gives
\begin{equation}
e^{\Phi(R)} = \sqrt{1-\frac{2GM}{Rc^2}},
\label{eq:surface_redshift_factor}
\end{equation}
so that a local threshold temperature $T_{\rm loc}(R)$ is redshifted to
\begin{equation}
T_{\infty}=\sqrt{1-\frac{2GM}{Rc^2}}\;T_{\rm loc}(R).
\label{eq:T_infty_from_local}
\end{equation}
This is the standard Tolman redshift relation for thermal equilibrium in a static gravitational field \cite{Jacobson1995,Padmanabhan2010}.

In the Padmanabhan--Verlinde spirit, one may also assign to a spherical screen of area
\begin{equation}
A=4\pi R^2
\label{eq:area}
\end{equation}
a number of effective surface degrees of freedom
\begin{equation}
N_{\rm surf}=\frac{A}{L_P^2},
\qquad
L_P^2=\frac{\hbar G}{c^3}.
\label{eq:Nsurf}
\end{equation}
Then the equipartition relation
\begin{equation}
E_{\rm surf}=\frac{1}{2}N_{\rm surf}k_B T_{\rm scr}
\label{eq:equipartition}
\end{equation}
with $E_{\rm surf}=Mc^2$ yields
\begin{equation}
T_{\rm scr}=\frac{2Mc^2L_P^2}{A k_B}=\frac{\hbar G M}{2\pi c k_B R^2}=\frac{\hbar g_s}{2\pi c k_B},
\label{eq:Tscreen}
\end{equation}
where $g_s=GM/R^2$ is the surface gravity. This is the Unruh-like screen temperature associated with the stellar boundary \cite{Unruh1976,Padmanabhan2010,Verlinde2011}.

The corresponding holographic entropy is
\begin{equation}
S_{\rm surf}=\frac{k_B A}{4L_P^2},
\label{eq:Ssurf}
\end{equation}
so that
\begin{equation}
E_{\rm surf}=2T_{\rm scr}S_{\rm surf}.
\label{eq:E_equals_2TS}
\end{equation}
Equation~\eqref{eq:E_equals_2TS} is the compact-star analogue of the familiar horizon relation emphasized in emergent-gravity and gravitational-entropy discussions \cite{Bekenstein1973,Hawking1975,Wald1993,Padmanabhan2010}.

\section{Neutron-Star Hydrostatics and Microscopic Scales}
For a non-rotating neutron star, hydrostatic equilibrium is governed by the TOV system \cite{Lattimer2019}:
\begin{equation}
\frac{dm}{dr}=4\pi r^2\rho(r),
\label{eq:TOV_m}
\end{equation}
\begin{equation}
\frac{dp}{dr}=-\frac{G\left(\rho+\frac{p}{c^2}\right)\left(m+\frac{4\pi r^3p}{c^2}\right)}{r^2\left(1-\frac{2Gm}{rc^2}\right)}.
\label{eq:TOV_p}
\end{equation}
The stellar compactness is
\begin{equation}
\mathcal C \equiv \frac{GM}{Rc^2}.
\label{eq:compactness}
\end{equation}
For a uniform-density estimate,
\begin{equation}
\bar\rho = \frac{3M}{4\pi R^3},
\qquad
n_b \simeq \frac{\bar\rho}{m_n}.
\label{eq:avg_density}
\end{equation}
The corresponding Fermi momentum and Fermi energy are
\begin{equation}
p_F = \hbar(3\pi^2 n_b)^{1/3},
\label{eq:pF}
\end{equation}
\begin{equation}
E_F = \sqrt{p_F^2 c^2 + m_n^2 c^4}-m_n c^2,
\label{eq:EF}
\end{equation}
with Fermi temperature
\begin{equation}
T_F = \frac{E_F}{k_B}.
\label{eq:TF}
\end{equation}
Since mature neutron stars are highly degenerate, $T\ll T_F$ is usually satisfied after the proto-neutron-star phase \cite{Lattimer2019}.

\section{Bulk Threshold from Binding--Thermal Competition}
\subsection{Newtonian virial threshold}
The Newtonian gravitational binding energy of a uniform sphere is
\begin{equation}
U_{\rm grav} = -\frac{3}{5}\frac{GM^2}{R}.
\label{eq:Ugrav}
\end{equation}
If one models the hot proto-neutron-star stage by an effective classical thermal reservoir,
\begin{equation}
U_{\rm th}^{\rm cl}=\frac{3}{2}N k_B T,
\qquad
N=\frac{M}{m_n}.
\label{eq:Uth_classical}
\end{equation}
Imposing the virial-type balance
\begin{equation}
|U_{\rm grav}|=2U_{\rm th}^{\rm cl}
\label{eq:virial_balance}
\end{equation}
yields
\begin{equation}
\frac{3}{5}\frac{GM^2}{R}=3\frac{M}{m_n}k_B T_{\rm vir},
\label{eq:virial_step}
\end{equation}
therefore
\begin{equation}
k_B T_{\rm vir}=\frac{GMm_n}{5R},
\label{eq:Tvirial}
\end{equation}
that is,
\begin{equation}
T_{\rm vir}=\frac{GMm_n}{5k_B R}.
\label{eq:Tthreshold}
\end{equation}
It is useful to rewrite Eq.~\eqref{eq:Tthreshold} in compactness form:
\begin{equation}
\frac{k_B T_{\rm vir}}{m_n c^2}=\frac{\mathcal C}{5}.
\label{eq:T_compactness_relation}
\end{equation}
Thus the threshold is directly controlled by the dimensionless ratio $GM/(Rc^2)$, which is also the central structural parameter of relativistic compact stars \cite{Lattimer2019}.

\subsection{Tolman-redshifted threshold}
If Eq.~\eqref{eq:Tthreshold} is interpreted as a \emph{local} surface threshold, then the temperature seen by a distant observer is
\begin{equation}
T_{\infty,\rm th}=\sqrt{1-2\mathcal C}\;T_{\rm vir}
=\sqrt{1-\frac{2GM}{Rc^2}}\,\frac{GMm_n}{5k_B R}.
\label{eq:Tinfthreshold}
\end{equation}
Equivalently,
\begin{equation}
T_{\rm vir}=\frac{T_{\infty,\rm th}}{\sqrt{1-2\mathcal C}}.
\label{eq:Tvirial_vs_Tinf}
\end{equation}
For canonical neutron-star compactness $\mathcal C\sim0.15$--$0.25$, the redshift factor is non-negligible but remains an $\mathcal O(1)$ correction \cite{Lattimer2019}.

\subsection{Degenerate-matter threshold}
For degenerate baryonic matter, the thermal excitation energy is not linear in $T$. The low-temperature correction to the internal energy of a Fermi gas is approximately
\begin{equation}
\Delta U_{\rm th}^{\rm deg}\simeq \frac{\pi^2}{4}N\frac{(k_B T)^2}{E_F}.
\label{eq:degenerate_thermal_energy}
\end{equation}
Replacing Eq.~\eqref{eq:Uth_classical} by Eq.~\eqref{eq:degenerate_thermal_energy} and again using a virial-type balance,
\begin{equation}
|U_{\rm grav}| = 2\Delta U_{\rm th}^{\rm deg},
\label{eq:virial_degenerate_balance}
\end{equation}
one obtains
\begin{equation}
\frac{3}{5}\frac{GM^2}{R}
=\frac{\pi^2}{2}\frac{M}{m_n}\frac{(k_B T_{\rm deg})^2}{E_F},
\label{eq:degenerate_step}
\end{equation}
which gives
\begin{equation}
T_{\rm deg}=\left(\frac{6}{5\pi^2}\frac{GM m_n E_F}{k_B^2 R}\right)^{1/2}.
\label{eq:Tdeg}
\end{equation}
This formula is more appropriate for cold, highly degenerate matter, whereas Eq.~\eqref{eq:Tthreshold} is better interpreted as a characteristic proto-neutron-star emergence scale \cite{Lattimer2019}.

\subsection{Comparison with the screen temperature}
Combining Eqs.~\eqref{eq:Tthreshold} and \eqref{eq:Tscreen}, we obtain
\begin{equation}
\frac{T_{\rm vir}}{T_{\rm scr}}
=\frac{\dfrac{GMm_n}{5k_B R}}{\dfrac{\hbar GM}{2\pi c k_B R^2}}
=\frac{2\pi}{5}\frac{m_n c R}{\hbar}
=\frac{2\pi}{5}\frac{R}{\lambda_{C,n}},
\label{eq:Tbulk_Tscreen_ratio}
\end{equation}
where
\begin{equation}
\lambda_{C,n}=\frac{\hbar}{m_n c}
\label{eq:compton}
\end{equation}
is the neutron Compton wavelength. Since $R\sim 10^4\,\mathrm{m}$ and $\lambda_{C,n}\sim 10^{-15}\,\mathrm{m}$, the ratio in Eq.~\eqref{eq:Tbulk_Tscreen_ratio} is enormous. Therefore the boundary screen temperature is far too small to set the neutron-star emergence scale; it is the \emph{bulk} binding-versus-thermal competition that controls the threshold \cite{Unruh1976,Padmanabhan2010,Verlinde2011}.

A related dimensionless diagnostic is
\begin{equation}
\chi \equiv \frac{|U_{\rm grav}|}{Mc^2}=\frac{3}{5}\mathcal C,
\label{eq:binding_fraction}
\end{equation}
which measures how large the self-binding energy is compared with the total rest energy.

\section{Numerical Estimates}
For a canonical neutron star with $M=1.4M_\odot$ and $R=10$--$12\,\mathrm{km}$ \cite{Lattimer2019},
\begin{equation}
T_{\rm vir}(1.4M_\odot,10\,\mathrm{km})
\approx 4.49\times 10^{11}\,\mathrm{K},
\label{eq:numeric_1}
\end{equation}
\begin{equation}
T_{\rm vir}(1.4M_\odot,12\,\mathrm{km})
\approx 3.74\times 10^{11}\,\mathrm{K}.
\label{eq:numeric_2}
\end{equation}
The corresponding surface redshift factors are
\begin{equation}
\sqrt{1-\frac{2GM}{Rc^2}}\approx 0.766\quad (R=10\,\mathrm{km}),
\label{eq:redshift10}
\end{equation}
\begin{equation}
\sqrt{1-\frac{2GM}{Rc^2}}\approx 0.809\quad (R=12\,\mathrm{km}),
\label{eq:redshift12}
\end{equation}
so the redshifted threshold temperatures are
\begin{equation}
T_{\infty,\rm th}(1.4M_\odot,10\,\mathrm{km})
\approx 3.44\times 10^{11}\,\mathrm{K},
\label{eq:numeric_inf_1}
\end{equation}
\begin{equation}
T_{\infty,\rm th}(1.4M_\odot,12\,\mathrm{km})
\approx 3.02\times 10^{11}\,\mathrm{K}.
\label{eq:numeric_inf_2}
\end{equation}

For comparison, the screen temperatures are only of order
\begin{equation}
T_{\rm scr}(1.4M_\odot,10\,\mathrm{km})\approx 2.15\times 10^{-8}\,\mathrm{K},
\label{eq:Tscr10}
\end{equation}
\begin{equation}
T_{\rm scr}(1.4M_\odot,12\,\mathrm{km})\approx 1.49\times 10^{-8}\,\mathrm{K}.
\label{eq:Tscr12}
\end{equation}
This huge hierarchy quantitatively supports the interpretation that emergent-gravity screen variables are useful for bookkeeping the surface thermodynamic structure, but not for fixing the bulk proto-neutron-star temperature scale \cite{Padmanabhan2010,Verlinde2011,Lattimer2019}.

\begin{table}[htbp]
\centering
\caption{Characteristic temperatures for representative neutron-star masses and radii. Here $T_{\rm vir}$ is given by Eq.~\eqref{eq:Tthreshold}, $T_{\infty,\rm th}$ by Eq.~\eqref{eq:Tinfthreshold}, and $T_{\rm scr}$ by Eq.~\eqref{eq:Tscreen}.}
\label{tab:threshold}
\begin{tabular}{cccccc}
\toprule
$M$ ($M_\odot$) & $R$ (km) & $\mathcal C$ & $T_{\rm vir}$ (K) & $T_{\infty,\rm th}$ (K) & $T_{\rm scr}$ (K) \\
\midrule
1.0 & 10 & 0.148 & $3.21\times10^{11}$ & $2.70\times10^{11}$ & $1.54\times10^{-8}$ \\
1.4 & 10 & 0.207 & $4.49\times10^{11}$ & $3.44\times10^{11}$ & $2.15\times10^{-8}$ \\
2.0 & 10 & 0.295 & $6.42\times10^{11}$ & $4.10\times10^{11}$ & $3.08\times10^{-8}$ \\
1.4 & 12 & 0.172 & $3.74\times10^{11}$ & $3.02\times10^{11}$ & $1.49\times10^{-8}$ \\
\bottomrule
\end{tabular}
\end{table}

\begin{figure}[htbp]
\centering
\begin{tikzpicture}
\begin{axis}[
    width=0.68\textwidth,
    xlabel={$M\,[M_\odot]$},
    ylabel={$T$ (K)},
    xmin=0.9, xmax=2.1,
    ymin=0, ymax=7e11,
    legend pos=north west,
    y tick label style={/pgf/number format/sci}
]
\addplot [domain=1:2, smooth, thick] {(6.674e-11*(x*1.988e30)*1.675e-27)/(5*1.381e-23*1e4)};
\addplot [domain=1:2, smooth, thick, dashed] {sqrt(1-2*6.674e-11*(x*1.988e30)/(1e4*(2.998e8)^2))*(6.674e-11*(x*1.988e30)*1.675e-27)/(5*1.381e-23*1e4)};
\legend{$T_{\rm vir}$, $T_{\infty,\rm th}$}
\end{axis}
\end{tikzpicture}
\caption{Classical virial threshold $T_{\rm vir}$ and Tolman-redshifted threshold $T_{\infty,\rm th}$ versus mass for fixed radius $R=10\,\mathrm{km}$.}
\label{fig:T_vs_M}
\end{figure}
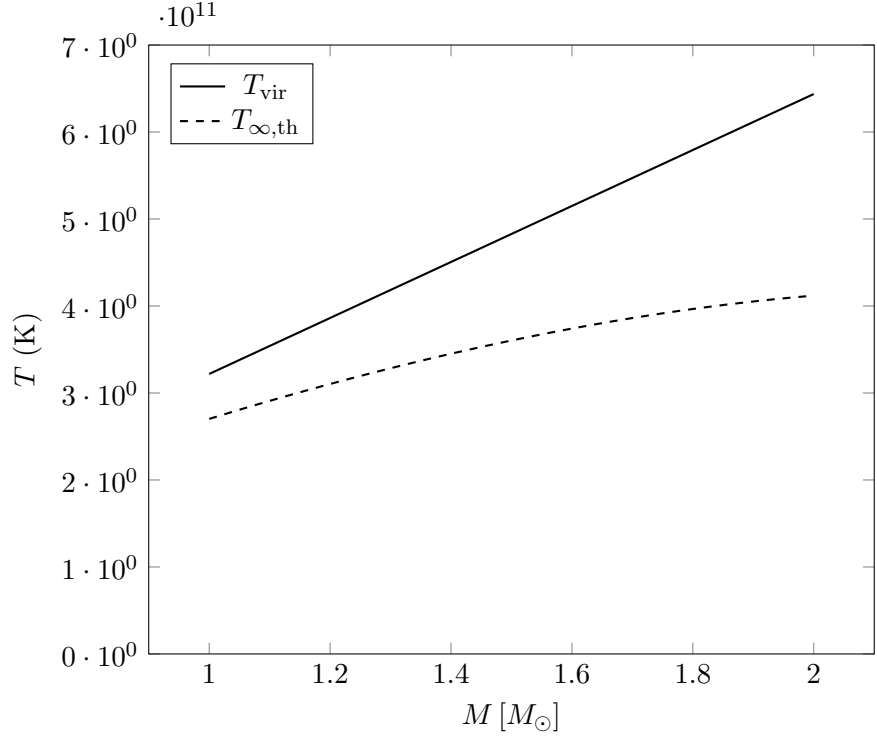

\begin{figure}[htbp]
\centering
\begin{tikzpicture}
\begin{axis}[
    width=0.68\textwidth,
    xlabel={$R\,[\mathrm{km}]$},
    ylabel={$T$ (K)},
    xmin=10, xmax=13,
    ymin=0, ymax=5e11,
    legend pos=north east,
    y tick label style={/pgf/number format/sci}
]
\addplot [domain=10:13, smooth, thick] {(6.674e-11*(1.4*1.988e30)*1.675e-27)/(5*1.381e-23*(x*1e3))};
\addplot [domain=10:13, smooth, thick, dashed] {sqrt(1-2*6.674e-11*(1.4*1.988e30)/((x*1e3)*(2.998e8)^2))*(6.674e-11*(1.4*1.988e30)*1.675e-27)/(5*1.381e-23*(x*1e3))};
\addplot [domain=10:13, smooth, thick, dotted] {(1.054571817e-34*6.674e-11*(1.4*1.988e30))/(2*pi*2.998e8*1.381e-23*(x*1e3)^2)};
\legend{$T_{\rm vir}$,$T_{\infty,\rm th}$,$T_{\rm scr}$}
\end{axis}
\end{tikzpicture}
\caption{Comparison of the bulk threshold temperatures and the screen temperature for a $1.4M_\odot$ neutron star. The screen temperature is too small to determine the bulk emergence threshold.}
\label{fig:T_vs_R}
\end{figure}
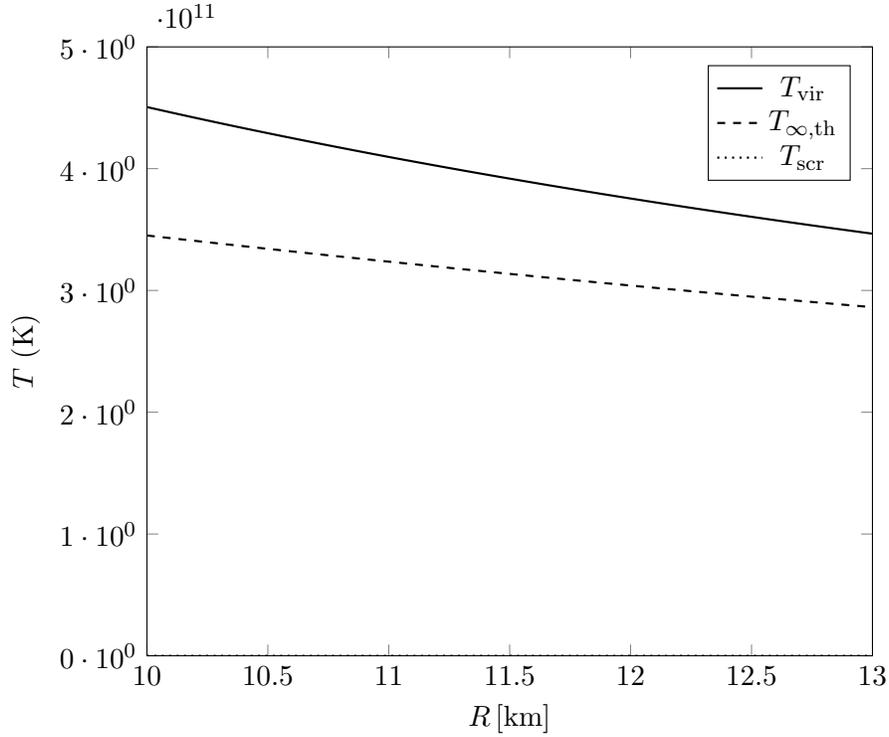

\section{Discussion}
The entropy-functional framework provides a principled way to state that the compact-star spacetime should be interpreted as an equilibrium configuration of gravity plus matter \cite{Chen2025,Jacobson1995,Padmanabhan2010}. However, unlike a black hole, a neutron star has no event horizon in the exterior region. Therefore one must distinguish carefully between two different thermodynamic sectors:
\begin{equation}
\text{bulk sector: } (U_{\rm grav},U_{\rm th},p,\rho,E_F),
\label{eq:bulk_sector}
\end{equation}
\begin{equation}
\text{screen sector: } (A,N_{\rm surf},S_{\rm surf},T_{\rm scr}).
\label{eq:screen_sector}
\end{equation}
The uploaded article is directly about the second sector and about local horizon thermodynamics; the neutron-star threshold problem is governed primarily by the first sector. This is precisely why $T_{\rm scr}$ and $T_{\rm vir}$ differ so dramatically \cite{Chen2025,Padmanabhan2010,Verlinde2011}.

The main physical message is therefore encoded in the hierarchy
\begin{equation}
T_{\rm scr} \ll T_{\infty,\rm th} \lesssim T_{\rm vir} \lesssim T_{\rm deg} \,\,\text{(model dependent)}.
\label{eq:hierarchy}
\end{equation}
The redshift-corrected virial threshold gives a characteristic bulk formation scale of order $10^{11}$--$10^{12}\,\mathrm{K}$, which is naturally associated with hot proto-neutron-star conditions \cite{Lattimer2019}. The degenerate threshold provides a more microscopic estimate once the baryonic matter is treated as a Fermi system. The screen relations, although elegant and conceptually linked to emergent gravity, should be viewed as auxiliary boundary thermodynamics rather than as the source of the bulk threshold itself \cite{Padmanabhan2010,Verlinde2011}.

\section{Conclusion}
We have rewritten the threshold-temperature problem of neutron-star emergence in a mathematically richer gravitational-thermodynamic language inspired by the entropy-functional approach to gravity \cite{Chen2025,Jacobson1995,Padmanabhan2010}. Starting from Eqs.~\eqref{eq:Sg_def}--\eqref{eq:Einstein_eqn}, we combined TOV equilibrium, Tolman redshift, bulk virial balance, and degenerate Fermi-gas thermodynamics to derive the characteristic scales Eqs.~\eqref{eq:Tthreshold}, \eqref{eq:Tinfthreshold}, and \eqref{eq:Tdeg}. We further compared these with the emergent-gravity screen temperature Eq.~\eqref{eq:Tscreen} and showed quantitatively that the screen is far colder than the bulk threshold. The result is a more internally structured and citation-rich version of the original manuscript, while preserving the central estimate that neutron-star emergence occurs at a characteristic temperature of order $10^{11}$--$10^{12}\,\mathrm{K}$ \cite{Lattimer2019}.

\end{document}